\pdfoutput=1
\documentclass[prl,aps,reprint,amsmath,amssymb,
superscriptaddress,nofootinbib]{revtex4-1}
\usepackage{epsfig}
\usepackage{graphicx}
\usepackage{cancel}
\usepackage{color}

\newcommand{\Ddots}{\hbox to 1em{.\hss.\hss.\hss}}

\begin{document}

\preprint{}

\title{Scalar BCJ Bootstrap}

\author{Taro V. Brown}

\affiliation{Center for Quantum Mathematics and Physics (QMAP), University of California, Davis, CA, USA}

\author{Karol Kampf}

\affiliation{Institute for Particle and Nuclear Physics, Charles University in Prague, Czech Republic}

\author{Umut Oktem}

\affiliation{Center for Quantum Mathematics and Physics (QMAP), University of California, Davis, CA, USA}

\author{Shruti Paranjape}

\affiliation{Center for Quantum Mathematics and Physics (QMAP), University of California, Davis, CA, USA}

\author{Jaroslav Trnka}

\affiliation{Center for Quantum Mathematics and Physics (QMAP), University of California, Davis, CA, USA}
\affiliation{Institute for Particle and Nuclear Physics, Charles University in Prague, Czech Republic}


\begin{abstract}
In this letter, we study tree-level scattering amplitudes of scalar particles in the context of effective field theories. We use tools similar to the soft bootstrap to build an ansatz for cyclically ordered amplitudes and impose the Bern-Carrasco-Johansson (BCJ) relations as a constraint. We obtain a set of BCJ-satisfying amplitudes as solutions to our procedure, which can be thought of as special higher-derivative corrections to SU(N) non-linear sigma model amplitudes satisfying BCJ relations to arbitrary multiplicity at leading order. The surprising outcome of our analysis is that BCJ conditions on higher-point amplitudes impose constraints on lower-point amplitudes, and they relate coefficients at different orders in the derivative expansion. This shows that BCJ conditions are much more restrictive than soft limit behavior, allowing only for a very small set of solutions.
\end{abstract}

\maketitle

\section{Introduction}

In recent years, the study of tree-level scattering amplitudes in effective field theories (EFT) has been a very active area of research, leading to very important insights ranging from the calculation of amplitudes in the Standard Model Effective Field Theory (SMEFT) to the intensive study of amplitudes in EFTs using modern methods such as soft bootstrap \cite{Kampf:2013vha,Cheung:2015cba, Cheung:2015ota, Cheung:2016drk,Elvang:2018dco}, scattering equations \cite{Cachazo:2014xea,Casali:2015vta} and color-kinematics duality \cite{Carrasco:2016ldy,Bern:2019prr, Chi:2021mio, Carrasco:2021ptp, Carrasco:2021ptp, Chen:2023dcx}.

A key model in these investigations is the SU(N) non-linear sigma model (NLSM) which describes the leading order dynamics of strong interactions at low energies. Pion scattering amplitudes vanish in the soft limit, i.e. possess an Adler zero, as a consequence of the shift symmetry of the Lagrangian. Higher-order corrections are organized in a derivative expansion in the context of Chiral Perturbation Theory ($\chi$PT) \cite{Weinberg:1978kz,Gasser:1983yg}. The Wilson coefficients in the $\chi$PT Lagrangian can be determined from experiments, but the theoretical derivation from the underlying QCD requires solving the theory at all scales which is incredibly hard. 

Tree-level amplitudes in QFT satisfy the important properties of \emph{locality} and \emph{unitarity}, i.e. all poles are located at $P^2=(p_{a_1}{+}{\dots}{+}p_{a_m})^2=0$ (for massless theories) and the amplitude factorizes on these poles into a product of two sub-amplitudes, schematically
\begin{equation}
A_n \xrightarrow[]{P^2=0} A_{n_1}\frac{1}{P^2}A_{n-n_1+2}\,.\label{fact}
\end{equation}
If the factorization conditions are enough to fix the amplitude uniquely, one can use (\ref{fact}) as input to reconstruct the $n$-point amplitude from lower-point amplitudes using recursion relations \cite{Britto:2004ap, Britto:2005fq}. In EFTs, this assumption is not satisfied because of the presence of contact terms. These have vanishing residues on all poles and hence are not constrained by factorization, so further properties of the amplitudes are needed to specify the model.

The bootstrap is a method that searches the space of all tree-level amplitudes for those that satisfy certain kinematic conditions. These are consequences of some underlying symmetry of the (a priori unknown) Lagrangian, but here we construct the amplitude from the bottom-up, without reference to Lagrangian operators. 
%
%
For scalar EFTs, we start with a generic rational function of variables $s_{ij}=(p_i+p_j)^2$, with poles at locations consistent with locality. We impose all possible residue equations (\ref{fact}) as constraints, followed by kinematic conditions to fix free parameters. A natural condition for low-energy EFTs is the soft limit constraint, i.e. for small momentum $p$ the amplitude behaves as
\begin{equation}
\lim_{p\rightarrow0} A_n = {\cal O}(p^\sigma)\,,\quad \mbox{for some integer $\sigma$.}\label{soft}
\end{equation}
Using this soft bootstrap, for $\sigma=1$ we have the standard Adler zero condition. If all coefficients are then fully specified, the amplitude is unique and can be reconstructed using recursion relations. This is the case for NLSM amplitudes \cite{Cheung:2015ota}. If some coefficients are left unfixed, there are more solutions which can be explained by the existence of multiple independent Lagrangians.

In this letter, we carry out a BCJ bootstrap by replacing the soft limit condition (\ref{soft}) by the Bern-Carrasco-Johansson (BCJ) relations
\begin{equation}
    \sum_{i=2}^{n{-}1} (s_{12}{+}{\dots}{+}s_{1i})A_n(2,{\dots},i,1,i{+}1,{\dots},n) = 0
\end{equation}
applied as a constraint to a local ansatz (together with the standard Kleiss-Kuijf (KK) relations \cite{Kleiss:1988ne}). The BCJ relations are famously satisfied by Yang-Mills amplitudes, as well as NLSM amplitudes, and play a crucial role in the double copy \cite{Bern:2019prr}. In particular, the double copy of a pair of Yang-Mills amplitudes gives rise to gravity amplitudes, while the double copy of NLSM amplitudes produces special Galileon amplitudes \cite{Cachazo:2014xea}. 

Our goal is to search for solutions to the BCJ relations in the space of ordered scalar amplitudes. As it turns out, the BCJ relations imply the Adler zero, and we naturally get a special selection of higher-derivative NLSM corrections. Surprisingly, we get constraints on the couplings not only at fixed multiplicity and derivative order but also between terms at different multiplicities and orders.

\medskip

\noindent {\bf Organization of the letter:} We begin with a review of NLSM amplitudes and the soft bootstrap in the context of NLSM corrections. We then classify all 4-point BCJ-satisfying amplitudes and formulate the BCJ bootstrap procedure at 6-point where we proceed up to order ${\cal O}(p^{18})$ with some surprising results. We do limited consistency checks for eight-point amplitudes and comment on the prospects of the existence of a ``BCJ Lagrangian''. We comment on the connection to the Z-theory amplitude, an important solution to the BCJ constraints \cite{Broedel:2013tta, Carrasco:2016ldy} that appears in the context of the CHY formula and double copy as the ``stringy part'' of open string amplitudes.

\section{Soft bootstrap for NLSM Amplitudes}

\noindent The tree-level amplitudes in the SU(N) non-linear sigma model can be decomposed into flavor-ordered sectors,
\begin{equation}
    {\cal A}^{\rm NLSM}_n = \sum_{\sigma} {\rm Tr}(T^{a_1}T^{a_2}{\dots}T^{a_n})\,A^{\rm NLSM}_n(1,2,{\dots},n) \,,
\end{equation}
where $A^{\rm NLSM}_n(1,2,{\dots},n)$ is a cyclically symmetric function,  $T^a$ are the generators of the SU(N) and we sum over all permutations $\sigma$ modulo cyclic ones. The only poles of $A^{\rm NLSM}_n$ are located at $P_{ij}^2 = (p_i{+}{\dots}{+}p_{j{-}1})^2 = 0$, and we use this fact as an input in the ansatz. 

At four-point, there are no factorization constraints and the size of the ansatz is directly equal to the number of independent four-point amplitudes. The cyclic symmetry implies that $A_4$ is symmetric in $s_{12}\leftrightarrow s_{23}$, and we can write the general form for the powercounting $A_4\sim s^m$ as
\begin{equation}
{\cal O}(p^{2m}):\quad A_4 \in \big\{u^{m-a}(s^a+t^a)\big\}\,,\label{4pt}
\end{equation}
where $a=0,2,{\dots},m$, $s=(p_1{+}p_2)^2$ and $t=(p_2{+}p_3)^2$. This means that we have $m$ independent four-point amplitudes at order ${\cal O}(p^{2m})$. At leading order, $A^{\rm NLSM}_n\sim \mathcal{O}(p^2)$. Using $m=1$ in \eqref{4pt} gives
\begin{equation}
A_4 = (p_1+p_3)^2 = u\,,
\end{equation}
the familiar four-pion NLSM amplitude. Power counting of terms in (\ref{4pt}) can also be seen in the context of the Lagrangian in a derivative expansion,
\begin{equation}
{\cal L}_{\chi PT} = {\cal L}_2 + {\cal L}_4 + {\cal L}_6 + \dots\,, \label{chpt}
\end{equation}
where the Lagrangian ${\cal L}_{2m}$ contains operators with $2m$ derivatives that are invariant under chiral symmetry. The usual building blocks are
\begin{align}
\label{eq:udef}
    u_\mu = i(u^\dagger \partial_\mu u{-}u\partial_\mu u^\dagger)\ \ \text{ where }\ u=\text{exp}\left(\frac{i \phi^aT^a}{F\sqrt{2}}\right),
\end{align}
$F$ is the pion decay constant. The corresponding scattering amplitudes are homogeneous functions of momenta that scale as $A_n\sim p^{2m}$. Each Lagrangian ${\cal L}_{2m}$ contains multiple independent terms (the number increases rapidly with $m$) and construction of all such Lagrangians has been an active area of research \cite{Bijnens:2001bb,Bijnens:2018lez}. 

Four-point amplitudes are special because the soft behavior is completely automatic for any non-constant function $F(s,t,u)$ because $s,t,u\rightarrow0$ when any $p_i\rightarrow0$. The first non-trivial constraints in the soft bootstrap occur at six-point. Here our ansatz consists of factorization terms that are determined by the four-point amplitudes and all possible contact terms, schematically
\begin{equation}
A_6=\sum_{\text{cyclic}}~~\raisebox{-6.5mm}{\includegraphics[trim={0cm 0cm 0cm 0cm},clip,scale=1]{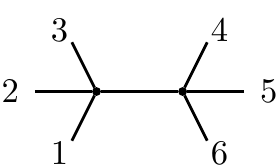}}~+~\raisebox{-8.9mm}{\includegraphics[trim={0cm 0cm 0cm 0cm},clip,scale=1]{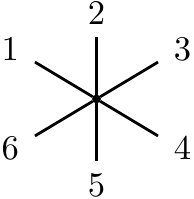}}\,.
\end{equation}
We could enlarge our ansatz to consider diagrams with more general kinematical invariants in the numerators, but using this ansatz is sufficient to guarantee correct factorization on all poles. On the other hand, soft limit behavior is not guaranteed. Additionally, there is always an ambiguity in the form of the 4-point amplitude (inserted into the vertices of the 6-point factorization diagram) as various equivalent 4-point forms do not agree on 6-point kinematics. It is convenient to choose the form written only in terms of external legs (not the internal leg $P$). The difference from any other form is just a contact term, which we add in the ansatz anyway.
%
%
\begin{equation}
 A_6^{\text{ans}}=\raisebox{-7.5mm}{\includegraphics[trim={0cm 0cm 0cm 0cm},clip,scale=0.8]{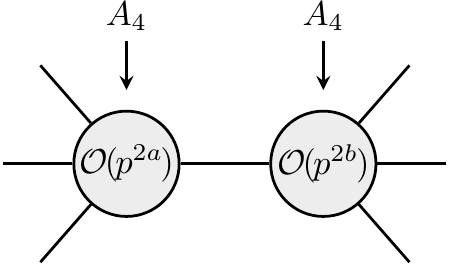}}+\raisebox{-8.5mm}{\includegraphics[trim={0cm 0cm 0cm 0cm},clip,scale=0.8]{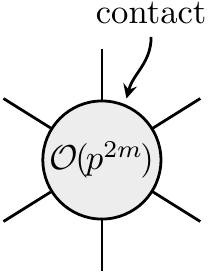}}\,.\label{Op6anForm}
\end{equation}
At a fixed derivative order ${\cal O}(p^{2m})$ the contact term involves $A_6\sim s^m$, while the factorization terms contain vertices $A_4^{(1)}\sim s^{a}$ and $A_4^{(2)}\sim s^{b}$ where $a{+}b=m{+}1$. For example at ${\cal O}(p^6)$ we have
\begin{equation}
\raisebox{-5.5mm}{\includegraphics[trim={0cm 0cm 0cm 0cm},clip,scale=0.8]{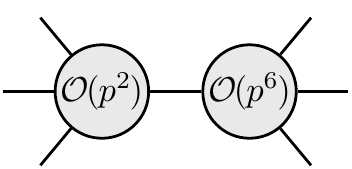}}+\raisebox{-5.5mm}{\includegraphics[trim={0cm 0cm 0cm 0cm},clip,scale=0.8]{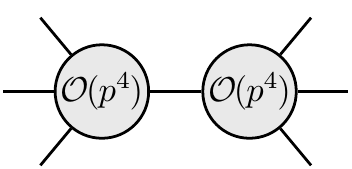}}+\hspace*{-0.05cm}\raisebox{-6.5mm}{\includegraphics[trim={0cm 0cm 0cm 0cm},clip,scale=0.8]{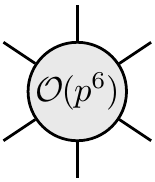}}\label{Op6an}\,.
\end{equation}
A special case is leading order NLSM i.e. $m{=}a{=}b{=}1$. Here the contact contribution is uniquely fixed by soft behavior,
\begin{equation}
A_6^{\rm NLSM} = \left(\frac{s_{13}s_{46}}{s_{123}}{+}\frac{s_{26}s_{35}}{s_{345}}{+}\frac{s_{15}s_{24}}{s_{234}}\right){-}\frac12\sum_{\rm cycl} s_{13}\,.
\end{equation}
In fact, all $n$-point amplitudes in the NLSM are uniquely fixed by soft recursion relations \cite{Cheung:2015cba, Cheung:2015ota, Cheung:2016drk} based on the 4-point amplitude. This uniqueness is also reflected in the fact that for massless pions, the leading order ${\cal O}(p^2)$ term in the $\chi$PT Lagrangian has only one term.

At general ${\cal O}(p^{2m})$ order, we have multiple solutions. The results at 6-point are summarized on the second line of Table \ref{tab4pt}. A similar analysis has been used to count independent terms in the $\chi$PT Lagrangian \cite{Dai:2020cpk}. To do this, one has to calculate higher-point amplitudes because at a given derivative order, some operators only contribute to eight- or higher-point amplitudes. 

Note two important features of this analysis: First, constraints in the ansatz are always imposed at a fixed ${\cal O}(p^{2m})$ order, e.g. the cyclic sum of the second type of diagram in the ${\cal O}(p^6)$ ansatz \eqref{Op6an} must satisfy the soft limit by itself. Though we still find solutions of the form
\begin{equation}
 A_6^{\text{ans}}=\raisebox{-6mm}{\includegraphics[trim={0cm 0cm 0cm 0cm},clip,scale=0.9]{p6-fac-2}}+\raisebox{-6.8mm}{\includegraphics[trim={0cm 0cm 0cm 0cm},clip,scale=0.9]{p6-contact}}\label{Op6an2}\,,
\end{equation}
in our analysis, the allowed contact term is very special and it must come from collapsing the propagator in the factorization diagram, i.e. it must have the form 
\begin{equation}
\label{eq:contactform}
A_6^{\rm contact} = \sum s_{ab}s_{cd}s_{ef}\,,
\end{equation}
where legs $a,b\in (1,2,3)$ and $c,d,e,f\in (4,5,6)$ (and cyclic shifts of external labels). This is indeed true. If this were not true, the coupling constants in the ${\cal O}(p^4)$ Lagrangian would be linked to the ${\cal O}(p^6)$ Lagrangian, which is not possible when the only constraint placed is chiral symmetry. Secondly, all constraints are placed at fixed multiplicity i.e. the 6-point soft bootstrap places no constraints on 4-point coefficients. 

\section{Four-point BCJ amplitudes}

Our goal is to use BCJ relations as a constraint on the ansatz rather than the soft limit of the amplitude. In principle, we can impose the Adler zero and BCJ relations independently, but in fact every solution to the BCJ conditions automatically has an Adler zero. This was suggested in \cite{CarrilloGonzalez:2019fzc} (see also \cite{Wei:2023iay}), and all checks performed in this paper confirm it (at 6-point up to ${\cal O}(p^{18})$). Therefore, the Adler zero condition is a weaker constraint. The 4-point BCJ relation can be written as 
\begin{equation}
sA_4(1,2,3,4) - uA_4(1,3,2,4) = 0\,.
\end{equation}
Together with cyclicity this implies that at ${\cal O}(p^{2m})$ order we can write the canonically ordered amplitude as
\begin{equation}
    A_4(1,2,3,4)\equiv A_4 = u\cdot F^{(2m{-}4)}(s,t,u)\,,
\end{equation}
where $F^{(2m{-}4)}$ is a symmetric polynomial of degree $m{-}2$ in $s,t,u$. A basis for all symmetric polynomials is
\begin{equation}
\label{eq:Fbasis}
    F^{(2m)}_{a,b}\in \Big\{(stu)^a(s^2+t^2+u^2)^b\Big\}\quad \mbox{for $3a+2b=m$.} 
\end{equation}
It is easy to see that the number of such terms is $[(m{+}2)/4]-[(m{+}2)/6]$ as previously reported \cite{Kampf:2021jvf, Carrasco:2019yyn}. The results are summarized in Table \ref{tab4pt} up to ${\cal O}(p^{18})$. 
\begin{table}[ht]
\centering
\caption{Soft and BCJ bootstrap results at 4-point}
\label{tab4pt}
\begin{tabular}{ |c||c|c|c|c|c|c|c|c|c| } 
 \hline
 ${\cal O}(p^\#)$ &  2 &  4 &  6 &  8 & 10 & 12 & 14 & 16 & 18   \\ \hline
Soft amplitudes & 1 & 2 & 3 & 4 & 5 & 6 & 7 & 8 & 9 \\ \hline
BCJ amplitudes & 1 & 0 & 1 & 1 & 1 & 1 & 2 & 1 & 2  \\
 \hline
\end{tabular}
\end{table}

Note that at ${\cal O}(p^4)$ there are no BCJ solutions, as the equation $3a+2b=1$ has no integer solutions. Additionally, up to ${\cal O}(p^{12})$ there is only one BCJ 4-point amplitude at each order, but starting ${\cal O}(p^{14})$ there are two (or more) independent solutions. Any BCJ-satisfying 4-point amplitude can hence be expanded in a basis,
\begin{equation}
\label{eq:Zth}
    A_4^{\rm BCJ} = \sum_{m,a,b} \alpha_{a,b}^{(2m)} \left(u F^{a,b}_{2m{-}4}\right)\,.
\end{equation}

One particularly special solution to BCJ constraints is the $Z$-theory amplitude, the ``stringy'' part of the Abelian open string amplitude \cite{Carrasco:2016ldy}. The derivative expansion above coincides with the $\alpha'$-expansion in this case. The 4-point $Z$-amplitude is written as a combination of $\Gamma$-functions, the low energy expansion yields
\begin{align}
\frac{1}{u}A_4^{\rm Z{-}theory} &= 1 + \frac{s^2{+}t^2{+}u^2}{96\pi^2F^6} + \frac{\zeta_3(stu)}{8\pi^6F^8} + \frac{(s^2{+}t^2{+}u^2)^2}{7680F^{10}\pi^4} \nonumber\\ 
&\hspace{-1.7cm}  + \frac{(stu)(s^2{+}t^2{+}u^2)(\zeta_2\zeta_3+2\zeta_5)}{128\pi^{10}F^{12}} \nonumber\\ 
&\hspace{-1.7cm} + \frac{51\zeta_6(s^2{+}t^2{+}u^2)^3+8(stu)^2(31\zeta_6{+}32\zeta_3^2)}{32768\pi^{12}F^{14}}+ \ldots,\label{Zth}
\end{align}
where we related $\alpha'$ and the pion decay constant as $F^2\sim(2\pi^2\alpha')^{-1}$, to align it with the expansion in $\chi$PT. Each term individually satisfies the 4-point BCJ relations and the full Z-theory amplitude is one particular combination of them. In particular, starting $\mathcal{O}(p^{14})$, there are contributions that are compatible with 4-point BCJ but do not appear in the Z-theory amplitude. Note that in our BCJ bootstrap analysis so far, the coefficients in (\ref{Zth}) are just arbitrary coefficients in front of terms which individually satisfy BCJ relations.

\section{BCJ as a bootstrap constraint}

We now proceed to discuss 6-point amplitudes. In the spirit of the soft bootstrap, we fix the derivative order ${\cal O}(p^{2m})$ and write a local ansatz. The factorization diagrams consist of products of 4-point BCJ amplitudes of a given order with unfixed coefficients, while the contact term is given by a general polynomial ansatz. We then impose the 6-point BCJ relation
\begin{align}
    &s_{12} A_6(123456) + (s_{12}{+}s_{23})A_6(132456)\nonumber\\
    &\hspace{0.5cm} - (s_{25}{+}s_{26})A_6(134256) - s_{26}A_6(134526) = 0 \,, \label{BCJ6pt}
\end{align}
which constrains coefficients in both the factorization terms and the contact term. Constructing the ansatz in this way guarantees the satisfaction of (\ref{BCJ6pt}) on all factorization channels but not for generic kinematics. 

The non-existence of 4-point ${\cal O}(p^4)$ BCJ amplitudes means that the corresponding factorization diagrams,
\begin{equation}
\raisebox{-5.7mm}{\includegraphics[trim={0cm 0cm 0cm 0cm},clip,scale=0.8]{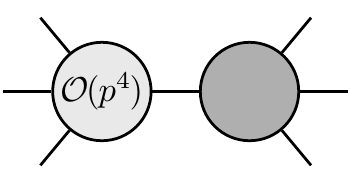}} \,\,\,\,\mbox{for any gray vertex,}
\end{equation}
are not in the ansatz, simplifying the analysis. For example, at ${\cal O}(p^6)$ there is only one factorization diagram consisting of the ${\cal O}(p^2)$ and ${\cal O}(p^{6})$ vertices along with the ${\cal O}(p^{6})$ contact term, i.e. (\ref{Op6an}) with the middle factorization diagram missing. One of the terms is
\begin{equation}
\begin{aligned}
   \raisebox{-7.2mm}{\includegraphics[trim={0cm 0cm 0cm 0cm},clip,scale=0.8]{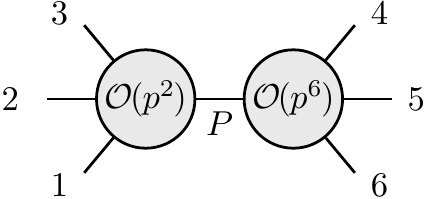}} =A_4^{(2)}(123)\frac{1}{P^2}A_4^{(6)}(456),\label{Op6-fac} 
\end{aligned}
\end{equation}
where we defined
\begin{align*}
A_4^{(2)}(123)&\equiv A_4^{(2)}(1,2,3,P) = s_{13},\\
A_4^{(6)}(456)&\equiv A_4^{(6)}(4,5,6,{-}P) = \alpha^{(6)}_{0,1} s_{46}(s_{45}^2{+}s_{46}^2{+}s_{56}^2),
\end{align*}
where $\alpha$'s are defined in \eqref{eq:Zth} and we set $\alpha^{(2)}_{0,0}=1$. Imposing the BCJ condition (\ref{BCJ6pt}) we fix the ${\cal O}(p^6)$ contact term uniquely and get the BCJ satisfying 6-point amplitude, see \cite{Carrasco:2016ldy} for explicit formula. The same procedure works at ${\cal O}(p^8)$ level. At ${\cal O}(p^{10})$, there is a new feature -- we get multiple factorization diagrams and some of them do not have a ${\cal O}(p^2)$ vertex,
\begin{equation}
 A_6^{\text{ans}}{=}\raisebox{-5.5mm}{\includegraphics[trim={0cm 0cm 0cm 0cm},clip,scale=0.8]{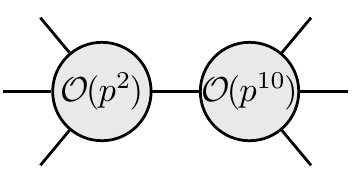}}{+}\raisebox{-5.5mm}{\includegraphics[trim={0cm 0cm 0cm 0cm},clip,scale=0.8]{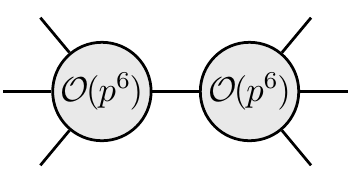}}{+}\hspace*{-0.05cm}\raisebox{-6.5mm}{\includegraphics[trim={0cm 0cm 0cm 0cm},clip,scale=0.8]{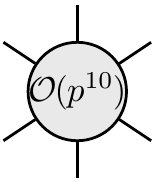}}.\label{Op10an}
\end{equation}
Imposing the BCJ condition we get two solutions: the first solution consists of the first factorization diagram and a contact term (i.e. the same form as (\ref{Op6anForm})) and again corresponds to some special term in the $\chi$PT expansion. The second solution is different:
\begin{equation}
 A_6^{\text{BCJ},2}=\raisebox{-5.5mm}{\includegraphics[trim={0cm 0cm 0cm 0cm},clip,scale=0.8]{p10-fac-2}}+\hspace*{-0.05cm}\raisebox{-6.5mm}{\includegraphics[trim={0cm 0cm 0cm 0cm},clip,scale=0.8]{p10-contact}}.\label{Op10an2}
\end{equation}
where the $s_{123}$ factorization diagram is now equal to
\begin{equation}
\left(\alpha_{0,1}^{(6)}\right)^2 \frac{s_{13}s_{46}(s_{12}^2{+}s_{13}^2{+}s_{23}^2)(s_{45}^2{+}s_{46}^2{+}s_{56}^2)}{s_{123}}\,.
\end{equation}
This is similar to the soft bootstrap at order ${\cal O}(p^6)$ where one factorization diagram had two ${\cal O}(p^4)$ vertices. In that case, the contact term had a special form \eqref{eq:contactform} and came from a collapsing propagator. Here the analysis reveals that in order to satisfy the BCJ relation in (\ref{Op10an2}), we need a genuine contact term not of the special form, and we get a unique solution. This means that BCJ relates coefficients in the ansatz for the ${\cal O}(p^{10})$ contact term in (\ref{Op10an}) to $(\alpha_{0,1}^{(6)})^2$. The Lagrangian that could generate such amplitudes is written schematically as
\begin{equation}
    {\cal L} = c_1 (\partial^6\phi^4) + c_2 (\partial^{10}\phi^6) + \dots
\end{equation}
where BCJ now relates coefficients $c_2$ and $c_1^2$. While this never happens in the context of $\chi$PT where the Lagrangian coefficients at different ${\cal O}(p^{2m})$ orders can not be related by soft physics, this phenomenon occurs in the study of enhanced soft limits leading to the Dirac-Born-Infeld and special Galileon theories \cite{Cheung:2016drk}. Here such conditions arise when we impose the BCJ relations. Continuing this analysis to the next order ${\cal O}(p^{12})$ we identify four solutions to the BCJ conditions, 
%
%
%
\begin{equation}
\begin{aligned}
A_6^{\text{BCJ},1}&=\raisebox{-5.4mm}{\includegraphics[trim={0cm 0cm 0cm 0cm},clip,scale=0.8]{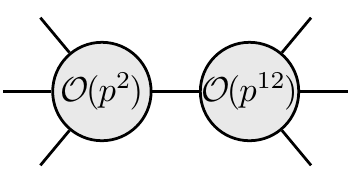}}+\raisebox{-6.5mm}{\includegraphics[trim={0cm 0cm 0cm 0cm},clip,scale=0.8]{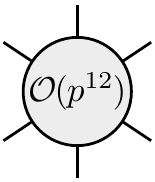}},\\
A_6^{\text{BCJ},2}&=\raisebox{-5.4mm}{\includegraphics[trim={0cm 0cm 0cm 0cm},clip,scale=0.8]{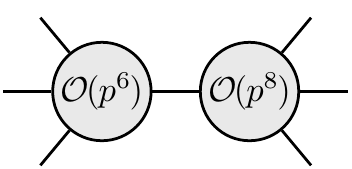}}+\raisebox{-6.5mm}{\includegraphics[trim={0cm 0cm 0cm 0cm},clip,scale=0.8]{bcj1-contact}},\\
A_6^{\text{BCJ},3}&=\raisebox{-6.5mm}{\includegraphics[trim={0cm 0cm 0cm 0cm},clip,scale=0.8]{bcj1-contact}},~~~A_6^{\text{BCJ},4}=\raisebox{-6.5mm}{\includegraphics[trim={0cm 0cm 0cm 0cm},clip,scale=0.8]{bcj1-contact}}.
\end{aligned}
\end{equation}
The first two solutions correspond to a factorization diagram, each completed by a particular ${\cal O}(p^{12})$ contact term, while the last two solutions are pure contact terms. Of course, any linear combination of BCJ amplitudes is a BCJ-satisfying amplitude. The results of the BCJ bootstrap are summarized in Table \ref{tab}, where the number of soft/BCJ amplitudes gives a total number of solutions, while the number of contact terms is a subset of that.
\begin{table}[ht]
\centering
\caption{Soft and BCJ bootstrap results at 6-point}
\label{tab}
\begin{tabular}{ |c|c|c|c|c|c|c|c|c|c|c| c|} 
 \hline
 ${\cal O}(p^\#)$ &  2 &  4 &  6 &  8 &  10 & 12 & 14   & 16 & 18  \\ \hline 
Soft amplitudes & 1 & 2 & 10 & 29 & 83 & 207 & 461  & 945 & 1819\\
 \hline
- Contact terms & 0 & 0 & 5 & 22 & 70 & 191 & 434 & 915 & 1772\\
 \hline 
 BCJ amplitudes &1 &0&1&1&2&4&7&16&36\\
 \hline
- Contact terms &0 &0&0&0&0&2&4&13&31\\
 \hline
\end{tabular}
\end{table}
%

\section{Four-point relations}

In principle, relations between coefficients at various ${\cal O}(p^{2m})$ come from UV physics and should not be accessible by any IR conditions -- certainly not soft limits. At order ${\cal O}(p^{14})$, we find first evidence of such relations arising from the BCJ bootstrap. Here something surprising happens: not all factorization diagrams with BCJ-satisfying 4-point amplitudes can be completed into BCJ-satisfying 6-point amplitudes by the addition of contact terms. The ansatz for the amplitude takes the form,
\begin{equation}
 \begin{aligned}
     A_6^{\text{ans}}=&~\raisebox{-5.5mm}{\includegraphics[trim={0cm 0cm 0cm 0cm},clip,scale=0.8]{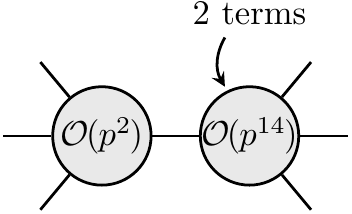}}+\raisebox{-5.5mm}{\includegraphics[trim={0cm 0cm 0cm 0cm},clip,scale=0.8]{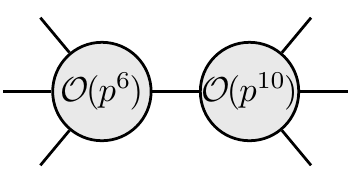}}\\
     &+\raisebox{-5.5mm}{\includegraphics[trim={0cm 0cm 0cm 0cm},clip,scale=0.8]{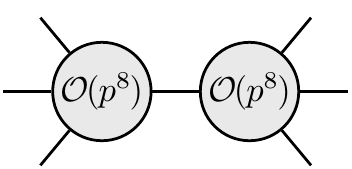}}+\raisebox{-6.5mm}{\includegraphics[trim={0cm 0cm 0cm 0cm},clip,scale=0.8]{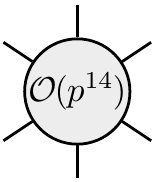}}.
 \end{aligned}\label{Op14an}
\end{equation}
Note that there are two different BCJ-satisfying ${\cal O}(p^{14})$ 4-point amplitudes, so the first factorization diagram represents two terms. Schematically, the ansatz is
\begin{align}
A_6^{\rm ans} &= \alpha^{(14)}_{2,0}({\dots}) + \alpha^{(14)}_{0,3}({\dots}) + \alpha^{(6)}_{0,1}\alpha^{(10)}_{0,2}({\dots}) \nonumber\\ & \hspace{2cm} + (\alpha^{(8)}_{1,0})^2({\dots}) + \sum_k \alpha^{\rm ct}_k({\dots})\,,
\end{align}
where $({\dots})$ stands for some kinematical expressions. Note that in this case there are four BCJ-satisfying contact terms which always appear as solutions to the BCJ conditions, even if we turn off the factorization terms. Modulo them, we find three BCJ-satisfying solutions. Naively one would expect four if all the 4-point BCJ-satisfying amplitudes get uplifted to 6-point. Instead, the 4-point coefficients must satisfy a constraint,
\begin{equation}
   \alpha_{2,0}^{(14)} - \frac{8}{3}\alpha_{0,3}^{(14)} - \frac{8}{3} \alpha^{(6)}_{0,1}\alpha^{(10)}_{0,2} - \frac12(\alpha^{(8)}_{1,0})^2 = 0\,,
\end{equation}
which allows us to solve for one of the parameters in terms of the others. It is interesting to see how the Z-theory amplitude satisfies this relation. The ${\cal O}(p^{14})$ coefficients have $\zeta_6$ and $\zeta_3^2$ parts while $\alpha^{(10)}_{0,2}\sim\zeta_4$, $\alpha^{(8)}_{1,0}\sim\zeta_3$ and 
$\alpha^{(6)}_{0,1}\sim\zeta_2$. The equality thus splits into an equation for the $\pi^6$ and $\zeta_3^2$ coefficients. 

At ${\cal O}(p^{16})$, the situation is similar to the ${\cal O}(p^{12})$ case. Applying the BCJ relations does not place any constraints on the 4-point coefficients. In other words, each factorization diagram individually leads to a BCJ-satisfying 6-point amplitude after the addition of appropriate contact terms. This seems to be related to the fact that there is only one ${\cal O}(p^{16})$ 4-point BCJ amplitude. 

At ${\cal O}(p^{18}
)$ order, the ansatz is 
\begin{align}
A_6^{\text{ans}}&=\raisebox{-5.5mm}{\includegraphics[trim={0cm 0cm 0cm 0cm},clip,scale=0.8]{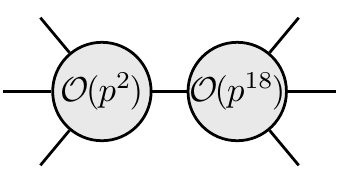}}+\raisebox{-5.5mm}{\includegraphics[trim={0cm 0cm 0cm 0cm},clip,scale=0.8]{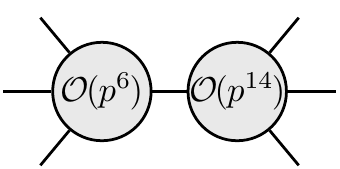}}  \label{Ans6pt18} \\
+&\raisebox{-5.5mm}{\includegraphics[trim={0cm 0cm 0cm 0cm},clip,scale=0.8]{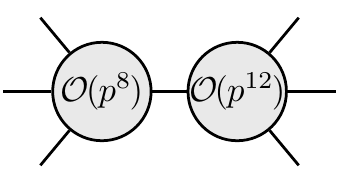}}+\raisebox{-5.5mm}{\includegraphics[trim={0cm 0cm 0cm 0cm},clip,scale=0.8]{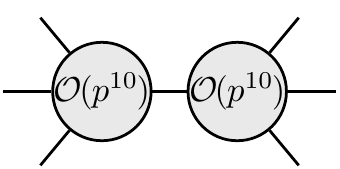}}+\raisebox{-6.5mm}{\includegraphics[trim={0cm 0cm 0cm 0cm},clip,scale=0.8]{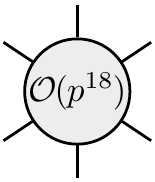}}\nonumber
\end{align}
Imposing the BCJ conditions, we find 31 contact terms (out of 2132 in the ansatz) satisfy our conditions, but not all factorization terms can be completed individually into BCJ-satisfying 6-point amplitudes. This is because we get one constraint on 4-point coupling constants,
\begin{align}
   & \alpha^{(18)}_{2,1} - 8\alpha^{(18)}_{0,4} + \alpha^{(6)}_{0,1}\alpha^{(14)}_{2,0} - 8 \alpha^{(6)}_{0,1}\alpha^{(14)}_{0,3} \nonumber\\ & \hspace{2.5cm} -\alpha^{(8)}_{1,0}\alpha^{(12)}_{1,1} -4 (\alpha^{(10)}_{0,2})^2 = 0\,, \label{Op18rel}
\end{align}
which has to be satisfied to get BCJ amplitudes. Hence in total at ${\cal O}(p^{18})$ we have 36 BCJ amplitudes, where we naively would have expected 37. The Z-theory 4-point amplitude again satisfies this relation in a non-trivial way, now dividing into three groups of terms proportional to $\zeta_8$, $\zeta_3\zeta_5$ and $\zeta_2\zeta_3^2$.

It is important to check if there are any new constraints coming from 8-point amplitudes. Generating these within the ansatz presents a challenge due to the extensive combinatorial complexity of the contact terms -- we were able to check up to ${\cal O}(p^{10})$. The general form of the ansatz for the 8-point amplitude is 
\begin{equation}
\begin{aligned}
A_8^{\text{ans}}=\raisebox{-6.5mm}{\includegraphics[trim={0cm 0cm 0cm 0cm},clip,scale=0.8]{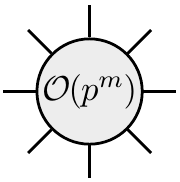}}\,+&\,\raisebox{-7.4mm}{\includegraphics[trim={0cm 0cm 0cm 0cm},clip,scale=0.8]{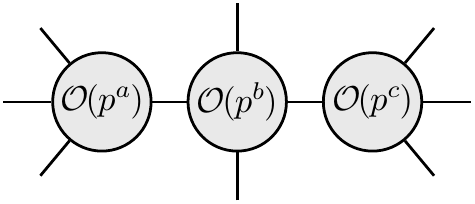}}\\
+\,\raisebox{-6.5mm}{\includegraphics[trim={0cm 0cm 0cm 0cm},clip,scale=0.8]{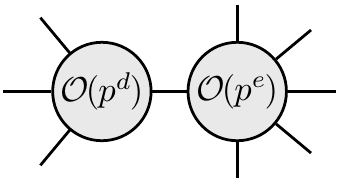}}&+\raisebox{-5.5mm}{\includegraphics[trim={0cm 0cm 0cm 0cm},clip,scale=0.8]{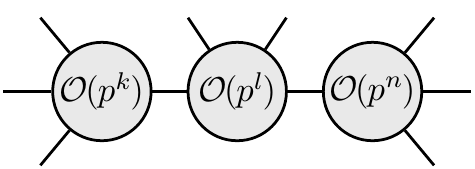}}.\label{Ans8pt}\\
\end{aligned}
\end{equation}
The power counting conditions are 
\begin{equation}
m=a{+}b{+}c{-}4 = d{+}e{-}2 =k{+}l{+}n{-}4\,.
\end{equation} 
We refer to the two types of diagrams as \emph{double-} and \emph{single-factorization} diagrams. Note that again we use BCJ-satisfying 4-point amplitudes in the factorization diagrams in the ansatz. For the 6-point vertex of the single-factorization diagram and for the 8-point contact term, we use a generic ansatz. In principle, on finding a solution, one should cross-check the solution for the 6-point vertex (appearing in the eight-point amplitude) with the solution for the BCJ-satisfying 6-point contact term, to find possible 6-point relations. Here, we do not do this and instead focus only on generating further conditions on the 4-point BCJ amplitudes. 

At ${\cal O}(p^6)$ and ${\cal O}(p^8)$ orders, nothing interesting happens: we find one BCJ 8-point amplitude at each order and no constraints on $\alpha^{(6)}_{0,1}$ and $\alpha^{(8)}_{1,0}$. The result takes the general form of (\ref{Ans8pt}), with one insertion of ${\cal O}(p^6)$ (resp. ${\cal O}(p^8)$) vertex and rest ${\cal O}(p^2)$ vertices. As there is no quadratic dependence on higher derivative terms, we could in principle expand both ${\cal O}(p^6)$ and ${\cal O}(p^8)$ BCJ solutions in the basis of terms in $\chi$PT. 

At ${\cal O}(p^{10})$ in the double-factorization terms in (\ref{Ans8pt}), we have either two ${\cal O}(p^6)$ vertices or one ${\cal O}(p^{10})$ vertex (the rest are ${\cal O}(p^2)$ vertices). The kinematical ansatz is
\begin{equation}
    A_8^{\rm ans} = \alpha^{(10)}_{0,2}({\dots}) + (\alpha^{(6)}_{0,1})^2({\dots}) + \sum_k \alpha_k^{\rm sf,ct}({\dots}) \,,
\end{equation}
where we grouped all single-factorization (sf) and contact (ct) terms in the last term, as we focus on the 4-point couplings here. Naively, we would expect two solutions which contain double-factorization terms if constants $\alpha^{(10)}_{0,2}$ and $\alpha^{(6)}_{0,1}$ are truly independent. Imposing the BCJ relations we find 21 solutions, but only one contains double factorization diagrams. Hence we get the following constraint on 4-point couplings,
\begin{equation}
\alpha^{(10)}_{0,2} = \frac65 (\alpha^{(6)}_{0,1})^2\,.
\end{equation}
This is of course satisfied by the Z-theory amplitude (\ref{Zth}). Note that the Abelian Z-theory amplitude \eqref{Zth} that we compare to, can be obtained by a sum over orderings of bi-colored Z-theory amplitudes. Recently, generalizations of such bi-colored Z-theory amplitudes were proposed \cite{Chen:2022shl, Chen:2023dcx}. The Abelianization of results in \cite{Chen:2022shl,Chen:2023dcx} can be achieved by a special selection of Wilson coefficients in our analysis. Thus, the analysis presented here is more general, leading to the natural question of whether there exists a different EFT description of bi-colored Z-theory.

\section{Outlook: towards a BCJ Lagrangian}

In this letter, we study the constraints placed on the coupling constants of higher-derivative scalar amplitudes by the BCJ relations. Starting with the BCJ-satisfying 4-point amplitude, we learn that when contributing to 6-point and 8-point BCJ amplitudes, the 4-point couplings must satisfy a number of constraints. BCJ relates terms of different derivative orders, which is very unlike how the Adler zero constrains amplitudes that give rise to $\chi$PT Lagrangians ${\cal L}_{\chi\text{PT}}$. Indeed the BCJ relations seem to interpolate between IR and UV physics, as they both imply the vanishing soft limit of amplitudes, as well as place constraints on couplings of various orders. We expect that demanding the BCJ relations are satisfied at higher-point will put even more constraints on lower-point amplitudes and their couplings. This raises an important question of whether the Z-theory amplitude is the only solution to this set of equations. 

The fact that BCJ relations constrain couplings at different multiplicity is linked to the question of the existence of ``BCJ Lagrangians'' which give rise to BCJ-satisfying amplitudes. We know one solution to this problem is the leading order 2-derivative NLSM Lagrangian. Additionally, we now know that in a putative higher-derivative Lagrangian, the couplings of various orders must be dependent -- we indeed found some of these dependencies, but more work is needed to establish if there exists a higher-derivative Lagrangian that satisfies the BCJ relations at arbitrary multiplicity. One possible solution to the problem is Z-theory whose Lagrangian, though unknown, is expected to exist. 

It is clear that such a BCJ Lagrangian is a special version of $\chi$PT Lagrangian with tuned Wilson coefficients as the Adler zero condition is included in the BCJ constrain. Since the $\chi$PT Lagrangian is known formally only up to a relatively low level ($O(p^8)$ in \cite{Bijnens:2018lez}) we can only explore a restricted case. General Lagrangian operators include covariant derivatives (e.g. $O(p^{10})$ can include $\nabla_\alpha u_\mu$ or $\nabla_\alpha \nabla_\beta u_\mu$ -- see \cite{Kampf:2021jvf} for details and notation). There is one class of terms we can write down, in principle up to all orders. These are terms starting with $n$ pions at $O(p^n)$, for generic dimension given as:
\begin{equation}
{\cal L}_{\chi\text{PT}}^n = \sum_{j=1}^{d_n} c_j \langle u_{\mu_{j_1}} \ldots u^{\mu_{j_1}} \ldots u_{\mu_{j_{n/2}}} \ldots u^{\mu_{j_{n/2}}} \rangle \,,
\end{equation}
where we sum over all chord diagrams ($d_n = 1, 2, 5, 17, \ldots$ as in \cite{Stoimenow}) and $u_\mu$ is defined in \eqref{eq:udef}. Interestingly, we find that BCJ imposes the following condition on the constants $c_j$:
\begin{equation}
\text{BCJ: }\qquad \sum_{j=1}^{d_n} c_j = 0 \,,\qquad \text{for }n>2\,.
\end{equation}
We have verified this conjecture up to $O(p^8)$. It might represent the structure of a necessary condition on the full higher-derivative Lagrangian and could be a hint in the search for a more general BCJ Lagrangian.

An even more ambitious goal is to see both the Adler zero and the BCJ relations arise from some underlying geometric structure in the framework of positive geometry \cite{Herrmann:2022nkh} such as the Amplituhedron \cite{Arkani-Hamed:2013jha,Arkani-Hamed:2017vfh} for planar ${\cal N}=4$ SYM amplitudes or the ABHY Associahedron for biadjoint $\phi^3$ amplitudes \cite{Arkani-Hamed:2017mur}.

\medskip

{\it Acknowledgements}: This work is supported by GA\-\v{C}R 21-26574S, MEYS LUAUS23126, DOE grant No. SC0009999 and the funds of the University of California.

\bibliography{mainbib}

\begin{thebibliography}{31}%
\makeatletter
\providecommand \@ifxundefined [1]{%
 \@ifx{#1\undefined}
}%
\providecommand \@ifnum [1]{%
 \ifnum #1\expandafter \@firstoftwo
 \else \expandafter \@secondoftwo
 \fi
}%
\providecommand \@ifx [1]{%
 \ifx #1\expandafter \@firstoftwo
 \else \expandafter \@secondoftwo
 \fi
}%
\providecommand \natexlab [1]{#1}%
\providecommand \enquote  [1]{``#1''}%
\providecommand \bibnamefont  [1]{#1}%
\providecommand \bibfnamefont [1]{#1}%
\providecommand \citenamefont [1]{#1}%
\providecommand \href@noop [0]{\@secondoftwo}%
\providecommand \href [0]{\begingroup \@sanitize@url \@href}%
\providecommand \@href[1]{\@@startlink{#1}\@@href}%
\providecommand \@@href[1]{\endgroup#1\@@endlink}%
\providecommand \@sanitize@url [0]{\catcode `\\12\catcode `\$12\catcode
  `\&12\catcode `\#12\catcode `\^12\catcode `\_12\catcode `\%12\relax}%
\providecommand \@@startlink[1]{}%
\providecommand \@@endlink[0]{}%
\providecommand \url  [0]{\begingroup\@sanitize@url \@url }%
\providecommand \@url [1]{\endgroup\@href {#1}{\urlprefix }}%
\providecommand \urlprefix  [0]{URL }%
\providecommand \Eprint [0]{\href }%
\providecommand \doibase [0]{http://dx.doi.org/}%
\providecommand \selectlanguage [0]{\@gobble}%
\providecommand \bibinfo  [0]{\@secondoftwo}%
\providecommand \bibfield  [0]{\@secondoftwo}%
\providecommand \translation [1]{[#1]}%
\providecommand \BibitemOpen [0]{}%
\providecommand \bibitemStop [0]{}%
\providecommand \bibitemNoStop [0]{.\EOS\space}%
\providecommand \EOS [0]{\spacefactor3000\relax}%
\providecommand \BibitemShut  [1]{\csname bibitem#1\endcsname}%
\let\auto@bib@innerbib\@empty
\bibitem [{\citenamefont {Kampf}\ \emph {et~al.}(2013)\citenamefont {Kampf},
  \citenamefont {Novotny},\ and\ \citenamefont {Trnka}}]{Kampf:2013vha}%
  \BibitemOpen
  \bibfield  {author} {\bibinfo {author} {\bibfnamefont {K.}~\bibnamefont
  {Kampf}}, \bibinfo {author} {\bibfnamefont {J.}~\bibnamefont {Novotny}}, \
  and\ \bibinfo {author} {\bibfnamefont {J.}~\bibnamefont {Trnka}},\ }\href
  {\doibase 10.1007/JHEP05(2013)032} {\bibfield  {journal} {\bibinfo  {journal}
  {JHEP}\ }\textbf {\bibinfo {volume} {05}},\ \bibinfo {pages} {032} (\bibinfo
  {year} {2013})}\BibitemShut {NoStop}%
\bibitem [{\citenamefont {Cheung}\ \emph {et~al.}(2015)\citenamefont {Cheung},
  \citenamefont {Shen},\ and\ \citenamefont {Trnka}}]{Cheung:2015cba}%
  \BibitemOpen
  \bibfield  {author} {\bibinfo {author} {\bibfnamefont {C.}~\bibnamefont
  {Cheung}}, \bibinfo {author} {\bibfnamefont {C.-H.}\ \bibnamefont {Shen}}, \
  and\ \bibinfo {author} {\bibfnamefont {J.}~\bibnamefont {Trnka}},\ }\href
  {\doibase 10.1007/JHEP06(2015)118} {\bibfield  {journal} {\bibinfo  {journal}
  {JHEP}\ }\textbf {\bibinfo {volume} {06}},\ \bibinfo {pages} {118} (\bibinfo
  {year} {2015})}\BibitemShut {NoStop}%
\bibitem [{\citenamefont {Cheung}\ \emph {et~al.}(2016)\citenamefont {Cheung},
  \citenamefont {Kampf}, \citenamefont {Novotny}, \citenamefont {Shen},\ and\
  \citenamefont {Trnka}}]{Cheung:2015ota}%
  \BibitemOpen
  \bibfield  {author} {\bibinfo {author} {\bibfnamefont {C.}~\bibnamefont
  {Cheung}}, \bibinfo {author} {\bibfnamefont {K.}~\bibnamefont {Kampf}},
  \bibinfo {author} {\bibfnamefont {J.}~\bibnamefont {Novotny}}, \bibinfo
  {author} {\bibfnamefont {C.-H.}\ \bibnamefont {Shen}}, \ and\ \bibinfo
  {author} {\bibfnamefont {J.}~\bibnamefont {Trnka}},\ }\href {\doibase
  10.1103/PhysRevLett.116.041601} {\bibfield  {journal} {\bibinfo  {journal}
  {Phys. Rev. Lett.}\ }\textbf {\bibinfo {volume} {116}},\ \bibinfo {pages}
  {041601} (\bibinfo {year} {2016})}\BibitemShut {NoStop}%
\bibitem [{\citenamefont {Cheung}\ \emph {et~al.}(2017)\citenamefont {Cheung},
  \citenamefont {Kampf}, \citenamefont {Novotny}, \citenamefont {Shen},\ and\
  \citenamefont {Trnka}}]{Cheung:2016drk}%
  \BibitemOpen
  \bibfield  {author} {\bibinfo {author} {\bibfnamefont {C.}~\bibnamefont
  {Cheung}}, \bibinfo {author} {\bibfnamefont {K.}~\bibnamefont {Kampf}},
  \bibinfo {author} {\bibfnamefont {J.}~\bibnamefont {Novotny}}, \bibinfo
  {author} {\bibfnamefont {C.-H.}\ \bibnamefont {Shen}}, \ and\ \bibinfo
  {author} {\bibfnamefont {J.}~\bibnamefont {Trnka}},\ }\href {\doibase
  10.1007/JHEP02(2017)020} {\bibfield  {journal} {\bibinfo  {journal} {JHEP}\
  }\textbf {\bibinfo {volume} {02}},\ \bibinfo {pages} {020} (\bibinfo {year}
  {2017})}\BibitemShut {NoStop}%
\bibitem [{\citenamefont {Elvang}\ \emph {et~al.}(2019)\citenamefont {Elvang},
  \citenamefont {Hadjiantonis}, \citenamefont {Jones},\ and\ \citenamefont
  {Paranjape}}]{Elvang:2018dco}%
  \BibitemOpen
  \bibfield  {author} {\bibinfo {author} {\bibfnamefont {H.}~\bibnamefont
  {Elvang}}, \bibinfo {author} {\bibfnamefont {M.}~\bibnamefont
  {Hadjiantonis}}, \bibinfo {author} {\bibfnamefont {C.~R.~T.}\ \bibnamefont
  {Jones}}, \ and\ \bibinfo {author} {\bibfnamefont {S.}~\bibnamefont
  {Paranjape}},\ }\href {\doibase 10.1007/JHEP01(2019)195} {\bibfield
  {journal} {\bibinfo  {journal} {JHEP}\ }\textbf {\bibinfo {volume} {01}},\
  \bibinfo {pages} {195} (\bibinfo {year} {2019})}\BibitemShut {NoStop}%
\bibitem [{\citenamefont {Cachazo}\ \emph {et~al.}(2015)\citenamefont
  {Cachazo}, \citenamefont {He},\ and\ \citenamefont {Yuan}}]{Cachazo:2014xea}%
  \BibitemOpen
  \bibfield  {author} {\bibinfo {author} {\bibfnamefont {F.}~\bibnamefont
  {Cachazo}}, \bibinfo {author} {\bibfnamefont {S.}~\bibnamefont {He}}, \ and\
  \bibinfo {author} {\bibfnamefont {E.~Y.}\ \bibnamefont {Yuan}},\ }\href
  {\doibase 10.1007/JHEP07(2015)149} {\bibfield  {journal} {\bibinfo  {journal}
  {JHEP}\ }\textbf {\bibinfo {volume} {07}},\ \bibinfo {pages} {149} (\bibinfo
  {year} {2015})}\BibitemShut {NoStop}%
\bibitem [{\citenamefont {Casali}\ \emph {et~al.}(2015)\citenamefont {Casali},
  \citenamefont {Geyer}, \citenamefont {Mason}, \citenamefont {Monteiro},\ and\
  \citenamefont {Roehrig}}]{Casali:2015vta}%
  \BibitemOpen
  \bibfield  {author} {\bibinfo {author} {\bibfnamefont {E.}~\bibnamefont
  {Casali}}, \bibinfo {author} {\bibfnamefont {Y.}~\bibnamefont {Geyer}},
  \bibinfo {author} {\bibfnamefont {L.}~\bibnamefont {Mason}}, \bibinfo
  {author} {\bibfnamefont {R.}~\bibnamefont {Monteiro}}, \ and\ \bibinfo
  {author} {\bibfnamefont {K.~A.}\ \bibnamefont {Roehrig}},\ }\href {\doibase
  10.1007/JHEP11(2015)038} {\bibfield  {journal} {\bibinfo  {journal} {JHEP}\
  }\textbf {\bibinfo {volume} {11}},\ \bibinfo {pages} {038} (\bibinfo {year}
  {2015})}\BibitemShut {NoStop}%
\bibitem [{\citenamefont {Carrasco}\ \emph {et~al.}(2017)\citenamefont
  {Carrasco}, \citenamefont {Mafra},\ and\ \citenamefont
  {Schlotterer}}]{Carrasco:2016ldy}%
  \BibitemOpen
  \bibfield  {author} {\bibinfo {author} {\bibfnamefont {J.~J.~M.}\
  \bibnamefont {Carrasco}}, \bibinfo {author} {\bibfnamefont {C.~R.}\
  \bibnamefont {Mafra}}, \ and\ \bibinfo {author} {\bibfnamefont
  {O.}~\bibnamefont {Schlotterer}},\ }\href {\doibase 10.1007/JHEP06(2017)093}
  {\bibfield  {journal} {\bibinfo  {journal} {JHEP}\ }\textbf {\bibinfo
  {volume} {06}},\ \bibinfo {pages} {093} (\bibinfo {year} {2017})}\BibitemShut
  {NoStop}%
\bibitem [{\citenamefont {Bern}\ \emph {et~al.}(2019)\citenamefont {Bern},
  \citenamefont {Carrasco}, \citenamefont {Chiodaroli}, \citenamefont
  {Johansson},\ and\ \citenamefont {Roiban}}]{Bern:2019prr}%
  \BibitemOpen
  \bibfield  {author} {\bibinfo {author} {\bibfnamefont {Z.}~\bibnamefont
  {Bern}}, \bibinfo {author} {\bibfnamefont {J.~J.}\ \bibnamefont {Carrasco}},
  \bibinfo {author} {\bibfnamefont {M.}~\bibnamefont {Chiodaroli}}, \bibinfo
  {author} {\bibfnamefont {H.}~\bibnamefont {Johansson}}, \ and\ \bibinfo
  {author} {\bibfnamefont {R.}~\bibnamefont {Roiban}},\ }\href@noop {} {\
  (\bibinfo {year} {2019})},\ \Eprint {http://arxiv.org/abs/1909.01358}
  {arXiv:1909.01358 [hep-th]} \BibitemShut {NoStop}%
\bibitem [{\citenamefont {Chi}\ \emph {et~al.}(2022)\citenamefont {Chi},
  \citenamefont {Elvang}, \citenamefont {Herderschee}, \citenamefont {Jones},\
  and\ \citenamefont {Paranjape}}]{Chi:2021mio}%
  \BibitemOpen
  \bibfield  {author} {\bibinfo {author} {\bibfnamefont {H.-H.}\ \bibnamefont
  {Chi}}, \bibinfo {author} {\bibfnamefont {H.}~\bibnamefont {Elvang}},
  \bibinfo {author} {\bibfnamefont {A.}~\bibnamefont {Herderschee}}, \bibinfo
  {author} {\bibfnamefont {C.~R.~T.}\ \bibnamefont {Jones}}, \ and\ \bibinfo
  {author} {\bibfnamefont {S.}~\bibnamefont {Paranjape}},\ }\href {\doibase
  10.1007/JHEP03(2022)077} {\bibfield  {journal} {\bibinfo  {journal} {JHEP}\
  }\textbf {\bibinfo {volume} {03}},\ \bibinfo {pages} {077} (\bibinfo {year}
  {2022})}\BibitemShut {NoStop}%
\bibitem [{\citenamefont {Carrasco}\ \emph {et~al.}(2021)\citenamefont
  {Carrasco}, \citenamefont {Rodina},\ and\ \citenamefont
  {Zekioglu}}]{Carrasco:2021ptp}%
  \BibitemOpen
  \bibfield  {author} {\bibinfo {author} {\bibfnamefont {J.~J.~M.}\
  \bibnamefont {Carrasco}}, \bibinfo {author} {\bibfnamefont {L.}~\bibnamefont
  {Rodina}}, \ and\ \bibinfo {author} {\bibfnamefont {S.}~\bibnamefont
  {Zekioglu}},\ }\href {\doibase 10.1007/JHEP06(2021)169} {\bibfield  {journal}
  {\bibinfo  {journal} {JHEP}\ }\textbf {\bibinfo {volume} {06}},\ \bibinfo
  {pages} {169} (\bibinfo {year} {2021})}\BibitemShut {NoStop}%
\bibitem [{\citenamefont {Chen}\ \emph {et~al.}(2023)\citenamefont {Chen},
  \citenamefont {Elvang},\ and\ \citenamefont {Herderschee}}]{Chen:2023dcx}%
  \BibitemOpen
  \bibfield  {author} {\bibinfo {author} {\bibfnamefont {A.~S.-K.}\
  \bibnamefont {Chen}}, \bibinfo {author} {\bibfnamefont {H.}~\bibnamefont
  {Elvang}}, \ and\ \bibinfo {author} {\bibfnamefont {A.}~\bibnamefont
  {Herderschee}},\ }\href@noop {} {\  (\bibinfo {year} {2023})},\ \Eprint
  {http://arxiv.org/abs/2302.04895} {arXiv:2302.04895 [hep-th]} \BibitemShut
  {NoStop}%
\bibitem [{\citenamefont {Weinberg}(1979)}]{Weinberg:1978kz}%
  \BibitemOpen
  \bibfield  {author} {\bibinfo {author} {\bibfnamefont {S.}~\bibnamefont
  {Weinberg}},\ }\href {\doibase 10.1016/0378-4371(79)90223-1} {\bibfield
  {journal} {\bibinfo  {journal} {Physica A}\ }\textbf {\bibinfo {volume}
  {96}},\ \bibinfo {pages} {327} (\bibinfo {year} {1979})}\BibitemShut
  {NoStop}%
\bibitem [{\citenamefont {Gasser}\ and\ \citenamefont
  {Leutwyler}(1984)}]{Gasser:1983yg}%
  \BibitemOpen
  \bibfield  {author} {\bibinfo {author} {\bibfnamefont {J.}~\bibnamefont
  {Gasser}}\ and\ \bibinfo {author} {\bibfnamefont {H.}~\bibnamefont
  {Leutwyler}},\ }\href {\doibase 10.1016/0003-4916(84)90242-2} {\bibfield
  {journal} {\bibinfo  {journal} {Annals Phys.}\ }\textbf {\bibinfo {volume}
  {158}},\ \bibinfo {pages} {142} (\bibinfo {year} {1984})}\BibitemShut
  {NoStop}%
\bibitem [{\citenamefont {Britto}\ \emph
  {et~al.}(2005{\natexlab{a}})\citenamefont {Britto}, \citenamefont {Cachazo},\
  and\ \citenamefont {Feng}}]{Britto:2004ap}%
  \BibitemOpen
  \bibfield  {author} {\bibinfo {author} {\bibfnamefont {R.}~\bibnamefont
  {Britto}}, \bibinfo {author} {\bibfnamefont {F.}~\bibnamefont {Cachazo}}, \
  and\ \bibinfo {author} {\bibfnamefont {B.}~\bibnamefont {Feng}},\ }\href
  {\doibase 10.1016/j.nuclphysb.2005.02.030} {\bibfield  {journal} {\bibinfo
  {journal} {Nucl. Phys. B}\ }\textbf {\bibinfo {volume} {715}},\ \bibinfo
  {pages} {499} (\bibinfo {year} {2005}{\natexlab{a}})}\BibitemShut {NoStop}%
\bibitem [{\citenamefont {Britto}\ \emph
  {et~al.}(2005{\natexlab{b}})\citenamefont {Britto}, \citenamefont {Cachazo},
  \citenamefont {Feng},\ and\ \citenamefont {Witten}}]{Britto:2005fq}%
  \BibitemOpen
  \bibfield  {author} {\bibinfo {author} {\bibfnamefont {R.}~\bibnamefont
  {Britto}}, \bibinfo {author} {\bibfnamefont {F.}~\bibnamefont {Cachazo}},
  \bibinfo {author} {\bibfnamefont {B.}~\bibnamefont {Feng}}, \ and\ \bibinfo
  {author} {\bibfnamefont {E.}~\bibnamefont {Witten}},\ }\href {\doibase
  10.1103/PhysRevLett.94.181602} {\bibfield  {journal} {\bibinfo  {journal}
  {Phys. Rev. Lett.}\ }\textbf {\bibinfo {volume} {94}},\ \bibinfo {pages}
  {181602} (\bibinfo {year} {2005}{\natexlab{b}})}\BibitemShut {NoStop}%
\bibitem [{\citenamefont {Kleiss}\ and\ \citenamefont
  {Kuijf}(1989)}]{Kleiss:1988ne}%
  \BibitemOpen
  \bibfield  {author} {\bibinfo {author} {\bibfnamefont {R.}~\bibnamefont
  {Kleiss}}\ and\ \bibinfo {author} {\bibfnamefont {H.}~\bibnamefont {Kuijf}},\
  }\href {\doibase 10.1016/0550-3213(89)90574-9} {\bibfield  {journal}
  {\bibinfo  {journal} {Nucl. Phys. B}\ }\textbf {\bibinfo {volume} {312}},\
  \bibinfo {pages} {616} (\bibinfo {year} {1989})}\BibitemShut {NoStop}%
\bibitem [{\citenamefont {Broedel}\ \emph {et~al.}(2013)\citenamefont
  {Broedel}, \citenamefont {Schlotterer},\ and\ \citenamefont
  {Stieberger}}]{Broedel:2013tta}%
  \BibitemOpen
  \bibfield  {author} {\bibinfo {author} {\bibfnamefont {J.}~\bibnamefont
  {Broedel}}, \bibinfo {author} {\bibfnamefont {O.}~\bibnamefont
  {Schlotterer}}, \ and\ \bibinfo {author} {\bibfnamefont {S.}~\bibnamefont
  {Stieberger}},\ }\href {\doibase 10.1002/prop.201300019} {\bibfield
  {journal} {\bibinfo  {journal} {Fortsch. Phys.}\ }\textbf {\bibinfo {volume}
  {61}},\ \bibinfo {pages} {812} (\bibinfo {year} {2013})}\BibitemShut
  {NoStop}%
\bibitem [{\citenamefont {Bijnens}\ \emph {et~al.}(2002)\citenamefont
  {Bijnens}, \citenamefont {Girlanda},\ and\ \citenamefont
  {Talavera}}]{Bijnens:2001bb}%
  \BibitemOpen
  \bibfield  {author} {\bibinfo {author} {\bibfnamefont {J.}~\bibnamefont
  {Bijnens}}, \bibinfo {author} {\bibfnamefont {L.}~\bibnamefont {Girlanda}}, \
  and\ \bibinfo {author} {\bibfnamefont {P.}~\bibnamefont {Talavera}},\ }\href
  {\doibase 10.1007/s100520100887} {\bibfield  {journal} {\bibinfo  {journal}
  {Eur. Phys. J. C}\ }\textbf {\bibinfo {volume} {23}},\ \bibinfo {pages} {539}
  (\bibinfo {year} {2002})}\BibitemShut {NoStop}%
\bibitem [{\citenamefont {Bijnens}\ \emph {et~al.}(2019)\citenamefont
  {Bijnens}, \citenamefont {Hermansson-Truedsson},\ and\ \citenamefont
  {Wang}}]{Bijnens:2018lez}%
  \BibitemOpen
  \bibfield  {author} {\bibinfo {author} {\bibfnamefont {J.}~\bibnamefont
  {Bijnens}}, \bibinfo {author} {\bibfnamefont {N.}~\bibnamefont
  {Hermansson-Truedsson}}, \ and\ \bibinfo {author} {\bibfnamefont
  {S.}~\bibnamefont {Wang}},\ }\href {\doibase 10.1007/JHEP01(2019)102}
  {\bibfield  {journal} {\bibinfo  {journal} {JHEP}\ }\textbf {\bibinfo
  {volume} {01}},\ \bibinfo {pages} {102} (\bibinfo {year} {2019})}\BibitemShut
  {NoStop}%
\bibitem [{\citenamefont {Dai}\ \emph {et~al.}(2020)\citenamefont {Dai},
  \citenamefont {Low}, \citenamefont {Mehen},\ and\ \citenamefont
  {Mohapatra}}]{Dai:2020cpk}%
  \BibitemOpen
  \bibfield  {author} {\bibinfo {author} {\bibfnamefont {L.}~\bibnamefont
  {Dai}}, \bibinfo {author} {\bibfnamefont {I.}~\bibnamefont {Low}}, \bibinfo
  {author} {\bibfnamefont {T.}~\bibnamefont {Mehen}}, \ and\ \bibinfo {author}
  {\bibfnamefont {A.}~\bibnamefont {Mohapatra}},\ }\href {\doibase
  10.1103/PhysRevD.102.116011} {\bibfield  {journal} {\bibinfo  {journal}
  {Phys. Rev. D}\ }\textbf {\bibinfo {volume} {102}},\ \bibinfo {pages}
  {116011} (\bibinfo {year} {2020})}\BibitemShut {NoStop}%
\bibitem [{\citenamefont {Carrillo~Gonz\'alez}\ \emph
  {et~al.}(2020)\citenamefont {Carrillo~Gonz\'alez}, \citenamefont {Penco},\
  and\ \citenamefont {Trodden}}]{CarrilloGonzalez:2019fzc}%
  \BibitemOpen
  \bibfield  {author} {\bibinfo {author} {\bibfnamefont {M.}~\bibnamefont
  {Carrillo~Gonz\'alez}}, \bibinfo {author} {\bibfnamefont {R.}~\bibnamefont
  {Penco}}, \ and\ \bibinfo {author} {\bibfnamefont {M.}~\bibnamefont
  {Trodden}},\ }\href {\doibase 10.1103/PhysRevD.102.105011} {\bibfield
  {journal} {\bibinfo  {journal} {Phys. Rev. D}\ }\textbf {\bibinfo {volume}
  {102}},\ \bibinfo {pages} {105011} (\bibinfo {year} {2020})}\BibitemShut
  {NoStop}%
\bibitem [{\citenamefont {Wei}\ and\ \citenamefont {Zhou}(2023)}]{Wei:2023iay}%
  \BibitemOpen
  \bibfield  {author} {\bibinfo {author} {\bibfnamefont {F.-S.}\ \bibnamefont
  {Wei}}\ and\ \bibinfo {author} {\bibfnamefont {K.}~\bibnamefont {Zhou}},\
  }\href@noop {} {\  (\bibinfo {year} {2023})},\ \Eprint
  {http://arxiv.org/abs/2305.04620} {arXiv:2305.04620 [hep-th]} \BibitemShut
  {NoStop}%
\bibitem [{\citenamefont {Kampf}(2021)}]{Kampf:2021jvf}%
  \BibitemOpen
  \bibfield  {author} {\bibinfo {author} {\bibfnamefont {K.}~\bibnamefont
  {Kampf}},\ }\href {\doibase 10.1007/JHEP12(2021)140} {\bibfield  {journal}
  {\bibinfo  {journal} {JHEP}\ }\textbf {\bibinfo {volume} {12}},\ \bibinfo
  {pages} {140} (\bibinfo {year} {2021})}\BibitemShut {NoStop}%
\bibitem [{\citenamefont {Carrasco}\ \emph {et~al.}(2020)\citenamefont
  {Carrasco}, \citenamefont {Rodina}, \citenamefont {Yin},\ and\ \citenamefont
  {Zekioglu}}]{Carrasco:2019yyn}%
  \BibitemOpen
  \bibfield  {author} {\bibinfo {author} {\bibfnamefont {J.~J.~M.}\
  \bibnamefont {Carrasco}}, \bibinfo {author} {\bibfnamefont {L.}~\bibnamefont
  {Rodina}}, \bibinfo {author} {\bibfnamefont {Z.}~\bibnamefont {Yin}}, \ and\
  \bibinfo {author} {\bibfnamefont {S.}~\bibnamefont {Zekioglu}},\ }\href
  {\doibase 10.1103/PhysRevLett.125.251602} {\bibfield  {journal} {\bibinfo
  {journal} {Phys. Rev. Lett.}\ }\textbf {\bibinfo {volume} {125}},\ \bibinfo
  {pages} {251602} (\bibinfo {year} {2020})}\BibitemShut {NoStop}%
\bibitem [{\citenamefont {Chen}\ \emph {et~al.}(2022)\citenamefont {Chen},
  \citenamefont {Elvang},\ and\ \citenamefont {Herderschee}}]{Chen:2022shl}%
  \BibitemOpen
  \bibfield  {author} {\bibinfo {author} {\bibfnamefont {A.~S.-K.}\
  \bibnamefont {Chen}}, \bibinfo {author} {\bibfnamefont {H.}~\bibnamefont
  {Elvang}}, \ and\ \bibinfo {author} {\bibfnamefont {A.}~\bibnamefont
  {Herderschee}},\ }\href@noop {} {\  (\bibinfo {year} {2022})},\ \Eprint
  {http://arxiv.org/abs/2212.13998} {arXiv:2212.13998 [hep-th]} \BibitemShut
  {NoStop}%
\bibitem [{\citenamefont {Stoimenow}(2000)}]{Stoimenow}%
  \BibitemOpen
  \bibfield  {author} {\bibinfo {author} {\bibfnamefont {A.}~\bibnamefont
  {Stoimenow}},\ }\href {\doibase 10.1016/S0012-365X(99)00347-7} {\bibfield
  {journal} {\bibinfo  {journal} {Discr. Math.}\ }\textbf {\bibinfo {volume}
  {218}},\ \bibinfo {pages} {209} (\bibinfo {year} {2000})}\BibitemShut
  {NoStop}%
\bibitem [{\citenamefont {Herrmann}\ and\ \citenamefont
  {Trnka}(2022)}]{Herrmann:2022nkh}%
  \BibitemOpen
  \bibfield  {author} {\bibinfo {author} {\bibfnamefont {E.}~\bibnamefont
  {Herrmann}}\ and\ \bibinfo {author} {\bibfnamefont {J.}~\bibnamefont
  {Trnka}},\ }\href {\doibase 10.1088/1751-8121/ac8709} {\bibfield  {journal}
  {\bibinfo  {journal} {J. Phys. A}\ }\textbf {\bibinfo {volume} {55}},\
  \bibinfo {pages} {443008} (\bibinfo {year} {2022})}\BibitemShut {NoStop}%
\bibitem [{\citenamefont {Arkani-Hamed}\ and\ \citenamefont
  {Trnka}(2014)}]{Arkani-Hamed:2013jha}%
  \BibitemOpen
  \bibfield  {author} {\bibinfo {author} {\bibfnamefont {N.}~\bibnamefont
  {Arkani-Hamed}}\ and\ \bibinfo {author} {\bibfnamefont {J.}~\bibnamefont
  {Trnka}},\ }\href {\doibase 10.1007/JHEP10(2014)030} {\bibfield  {journal}
  {\bibinfo  {journal} {JHEP}\ }\textbf {\bibinfo {volume} {10}},\ \bibinfo
  {pages} {030} (\bibinfo {year} {2014})}\BibitemShut {NoStop}%
\bibitem [{\citenamefont {Arkani-Hamed}\ \emph
  {et~al.}(2018{\natexlab{a}})\citenamefont {Arkani-Hamed}, \citenamefont
  {Thomas},\ and\ \citenamefont {Trnka}}]{Arkani-Hamed:2017vfh}%
  \BibitemOpen
  \bibfield  {author} {\bibinfo {author} {\bibfnamefont {N.}~\bibnamefont
  {Arkani-Hamed}}, \bibinfo {author} {\bibfnamefont {H.}~\bibnamefont
  {Thomas}}, \ and\ \bibinfo {author} {\bibfnamefont {J.}~\bibnamefont
  {Trnka}},\ }\href {\doibase 10.1007/JHEP01(2018)016} {\bibfield  {journal}
  {\bibinfo  {journal} {JHEP}\ }\textbf {\bibinfo {volume} {01}},\ \bibinfo
  {pages} {016} (\bibinfo {year} {2018}{\natexlab{a}})}\BibitemShut {NoStop}%
\bibitem [{\citenamefont {Arkani-Hamed}\ \emph
  {et~al.}(2018{\natexlab{b}})\citenamefont {Arkani-Hamed}, \citenamefont
  {Bai}, \citenamefont {He},\ and\ \citenamefont {Yan}}]{Arkani-Hamed:2017mur}%
  \BibitemOpen
  \bibfield  {author} {\bibinfo {author} {\bibfnamefont {N.}~\bibnamefont
  {Arkani-Hamed}}, \bibinfo {author} {\bibfnamefont {Y.}~\bibnamefont {Bai}},
  \bibinfo {author} {\bibfnamefont {S.}~\bibnamefont {He}}, \ and\ \bibinfo
  {author} {\bibfnamefont {G.}~\bibnamefont {Yan}},\ }\href {\doibase
  10.1007/JHEP05(2018)096} {\bibfield  {journal} {\bibinfo  {journal} {JHEP}\
  }\textbf {\bibinfo {volume} {05}},\ \bibinfo {pages} {096} (\bibinfo {year}
  {2018}{\natexlab{b}})}\BibitemShut {NoStop}%
\end{thebibliography}%


%
\bibliographystyle{apsrev4-1}

\end{document}